\documentclass{article}

\usepackage[final]{neurips_2023_ml4ps}

\usepackage[utf8]{inputenc} 
\usepackage[T1]{fontenc}    
\usepackage{hyperref}       
\usepackage{url}            
\usepackage{booktabs}       
\usepackage{amsfonts}       
\usepackage{nicefrac}       
\usepackage{microtype}      
\usepackage{xcolor}         
\usepackage{graphicx}
\usepackage{setspace}
\usepackage{cprotect}

\title{Latent space representations of cosmological fields}

\author{%
  Sambatra Andrianomena \\
  South African Radio Astronomy Observatory\\
  Department of Physics \& Astronomy, \\ 
  University of the Western Cape\\
  \texttt{andrianomena@gmail.com} \\
  \And
  Sultan Hassan~\thanks{NASA Hubble Fellow}\\
  New York University\\
  Flatiron Institute \\
  University of the Western Cape\\
  \texttt{sultan.hassan@nyu.edu} \\
}

\begin{document}

\maketitle

\begin{abstract}
We investigate the possibility of learning the representations of cosmological multifield dataset from the CAMELS project. We train a very deep variational encoder on images which comprise three channels, namely gas density (Mgas), neutral hydrogen density (HI), and magnetic field amplitudes (B). The clustering of the images in feature space with respect to some cosmological/astrophysical parameters (e.g. $\Omega_{\rm m}$) suggests that the generative model has learned latent space representations of the high dimensional inputs. We assess the quality of the latent codes by conducting a linear test on the extracted features, and find that a single dense layer is capable of recovering some of the parameters to a promising level of accuracy, especially the matter density whose prediction corresponds to a coefficient of determination $R^{2}$ = 0.93. Furthermore, results show that the generative model is able to produce images that exhibit statistical properties which are consistent with those of the training data, down to scales of $k\sim 4h/{\rm Mpc}.$
\end{abstract}

\section{Introduction}
The data deluge which will be coming from current and different surveys at different wavelengths, e.g. DESI \citep{aghamousa2016desi}, SKA \citep{dewdney2009square, weltman2020fundamental} and HIRAX \citep{newburgh2016hirax}, will improve our understanding of the large-scale structure. In order to uncover patterns and anomalies in the data from those surveys, advanced techniques will be required. Deep networks that are considered for parameter inference at the field level can help extract the salient features from high dimensional data (e.g. 2D images). Although cosmological simulations are  useful for training those convolutional neural network (CNN) based methods to recover the underlying cosmology and astrophysics \citep{gupta2018non, ribli2019improved, fluri2019cosmological, villaescusa2021multifield, andrianomena2022predictive}, extracting the relevant features from noisy data in real world scenario still proves to be challenging for those networks. One way to tackle that issue is to train the networks with noisy mock data \citep[e.g.][]{hassan2020constraining}, making them more robust when predicting inputs in the presence noise. However, systematics from real surveys can be hard to model, which still poses a problem. Another approach, which is less sensitive to complex noise, is to learn the representations of the high dimensional data. In general, representation learning which consists of extracting the salient features from the inputs does not rely on the labels of the latter and is suitable for discovering patterns and anomalies. Self-supervised learning \citep{chen2020simple, grill2020bootstrap, he2020momentum, chen2021exploring} has been proven to be able to successfully extract useful representations of images (e.g. CIFAR10). Our main objective in this work is to learn the latent codes of the multifield images which are composed of three 2D maps (channels) of different physical fields. This is crucial in preparation for analyzing and understanding the unlabelled data from different cosmological surveys. In contrast with self-supervised learning based approaches, our method -- a variant of variational autoencoder (VAE) -- is simultaneously trained to extract important information from the input, and generate \textit{new} images that have both the quality and statistical properties that are comparable to the those of the training data. Previous works also used VAE models to analyze cosmological data \citep[e.g.][]{troster2019painting, yi2020cosmo, thorne2021generative, horowitz2022hyphy}.

\section{Method}
\label{method}
\subsection{Data}\label{data}
We consider the public CAMELS Multifields Dataset (CMD) \citep{CMD} which was obtained from running state-of-the art hydrodynamic simulation IllustrisTNG from the CAMELS Project \citep{villaescusa2021camels} . In CMD, there are 13 different fields representing a region of $25\times25\times5~(h^{-1}{\rm Mpc})^3$ at $z=0$. In this proof of concept, each instance in our training/testing dataset comprises three channels corresponding to gas density (Mgas), neutral hydrogen (HI) and magnetic fields magnitude (B). The three different fields (or channels) in each example describe the same physical region with the same underlying cosmological $\Omega_{\rm m}$ and $\sigma_{8}$ and astrophysical parameters ($A_{\rm SN1}$, $A_{\rm SN2}$, $A_{\rm AGN1}$, $A_{\rm AGN2}$) characterizing the supernova and active galactic nuclei feedbacks. The training, validation and test sets contain 13,000, 1,000 and 1,000 maps of $256\times256$ pixels$^2$ respectively.

\subsection{Algorithm}
In our analyses, we consider the variational autoencoder  prescribed in \cite{child2020very}. VAE \citep{kingma2013auto} comprises an encoder  $q_{\phi}(z|x)$, a decoder $p_{\theta}(x|z)$ and a prior $p_{\theta}(z)$. The training of the two networks $\phi$ and $\theta$ consists of maximizing the evidence lower bound (ELBO) \citep{kingma2013auto} 
\begin{equation}
    \mathbb{ELBO} = E_{z\sim q_{\phi}(z|x)}{\rm log}p_{\theta}(x|z) - D_{KL}(q_{\phi}(z|x)|| p_{\theta}(z)),
\end{equation}
where the first term is the reconstruction loss, 
whereas the second term is the Kullback-Leibler divergence, measuring the dissimilarity between the approximate posterior $q_{\phi}(z|x)$ and the prior $p_{\theta}(z)$. 
\cite{child2020very} introduced a hierarchical VAE, which is composed of many stochastic layers of latent variables, such that both the approximate posterior and the prior, which generate the latent variables, are more expressive. The latent variables which are conditionally dependent on each other (like in autoregressive models) have different resolutions according to
\begin{eqnarray}\label{conditionals}
    p_{\theta}(z) = p_{\theta}(z_{0})\prod_{k = 1}^{N}p_{\theta}(z_{k}|z_{k-1}), \nonumber\\
    q_{\phi}(z|x) = q_{\phi}(z_{0}|x)\prod_{k = 1}^{N}q_{\phi}(z_{k}|z_{k-1}, x),
\end{eqnarray}
where $N$ is the number of layers, and $q_{\phi}(\cdot)$ and $p_{\theta}(\cdot)$ are diagonal Gaussian distributions. It is worth noting that, unless otherwise stated, the learned representations (or latent codes) in our analyses denote the latent variables with the lowest resolutions (vector of length 256). The encoder comprises many stages of stacked residual blocks and the downsampling after the last residual block in one stage is achieved with an average pooling. A residual block has 4 convolutional layers. A top-down block uses one residual block for the prior, one residual block for the posterior, one convolutional layer for the latent variable and a last residual block at its ouput. The decoder has the same number of stages as the encoder, and each of them is composed of stacked top-down blocks. The upsampling after the last top-down block at a given stage is done by using nearest neighor (see the schematic diagram in Figure 3 of \cite{child2020very}). During training, the conditionals in equation \ref{conditionals} are achieved by mixing (via concatenation) the output of the last residual block at a given stage of the encoder with the inputs of all top-down blocks at the corresponding stage of the decoder. We consider 6 stages with (3, 3, 2, 2, 2, 2) number of residual blocks for the encoder, and (3, 3, 2, 2, 2, 2) number of top-down blocks for the decoder. We train the model for 300 epochs using RMSprop optimizer with learning rate = 0.00002, momentum = 0.9 and weight decay = $10^{-4}$, in batches of 4 instances on a NVIDIA GeForce GTX 1080 Ti. Each epoch takes about 26 minutes.
\section{Results}\label{results}

\begin{figure}
  \centering
  \includegraphics[width=0.7\columnwidth]{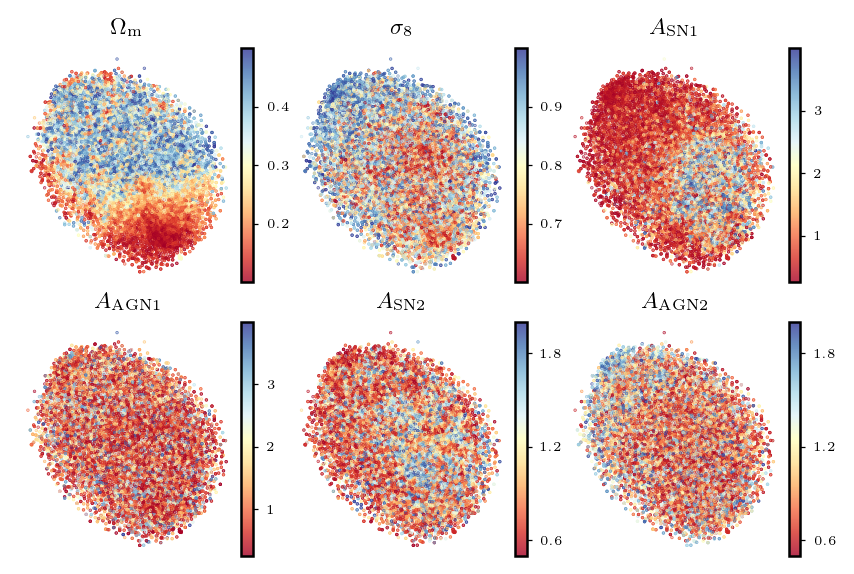}
  \cprotect\caption{Visualization of the learned representations using t-SNE. The color coding in each panel denotes the parameter values.}
  \label{fig:representation}
\end{figure}
\begin{figure}
  \centering
  \includegraphics[width=0.6\columnwidth]{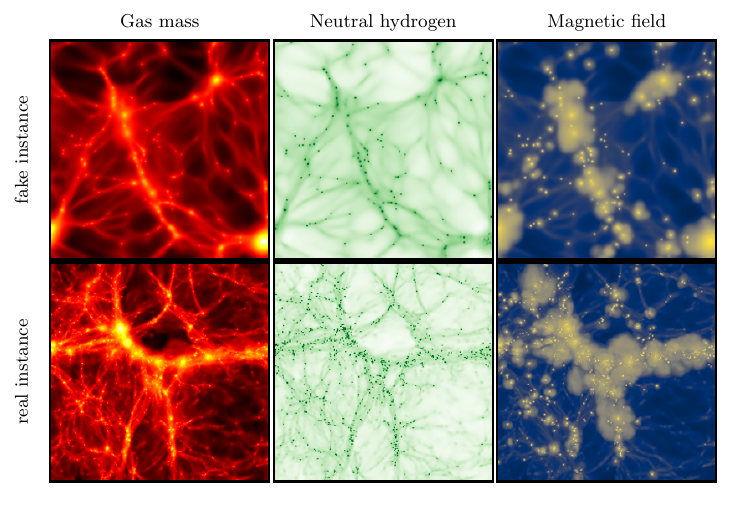}
  \cprotect\caption{Comparing field images of a new instance generated by the trained decoder on the top row with those of a real instance at the bottom row from the test set. Each field of both real and fake instances is shown in the same column.}
  \label{fig:fake images}
\end{figure}
\begin{figure}
  \centering
  \includegraphics[width=0.7\columnwidth]{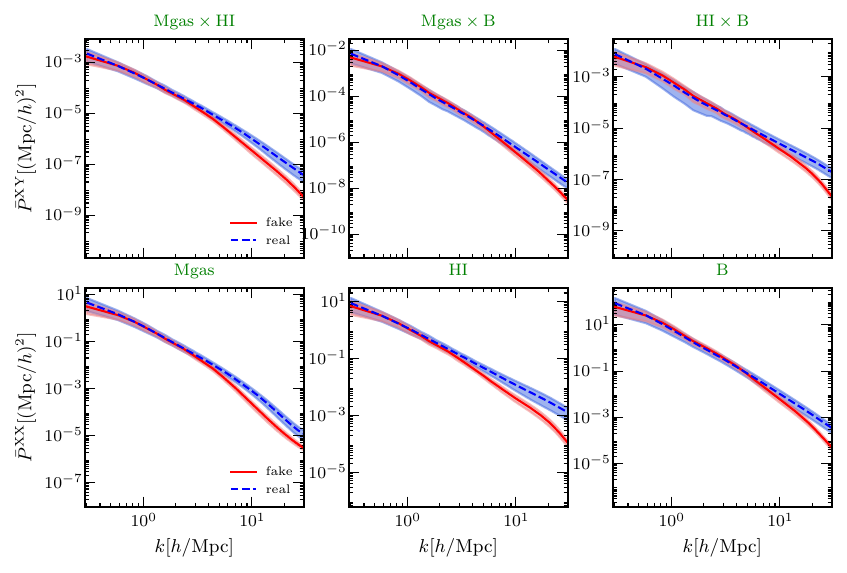}
  \cprotect\caption{Top row: average cross-power spectrum between the fields for both the test images (dashed blue) and generated images (solid red) as a function of wavenumber $k$. Bottom row: average auto-power spectrum of each field in the test images (dashed blue) and in the fake images (solid red). In all cases, the shaded area denotes the standard deviation of a power spectrum, blue and red for the real and fake respectively.}
  \label{fig:power spectra}
\end{figure}

To investigate the learned latent codes during training, the whole dataset is fed into the encoder in order to extract the features which consist of vectors of length 256. For visualization, the features are further projected into a two dimensional subspace using a dimensionality reduction method, t-distributed stochastic neighbor embedding (t-SNE). Figure \ref{fig:representation} shows the clustering of the data points in 2D space. Each point in each panel is color-coded according to the value of the underlying parameter, and each panel presents the clustering with respect to one parameter, e.g. $\Omega_{\rm m}$. Overall, the clustering of the latent representations is visible with respect to the cosmological parameters ($\Omega_{\rm m}$, $\sigma_{8}$) and the supernova feedback parameters ($A_{\rm SN1}$, $A_{\rm SN2}$), although the trend related to the AGN feedback parameters is very weak or non-existent. In other words, feature vectors associated with roughly the same value of any of those four parameters tend to be clumped together in the embedding space. This is indicative of the ability of the generative model to learn the latent codes of the inputs.

For the linear test, the new dataset composed of the features extracted (vector of length 256) from the entire dataset is similarily split into training, validation and testing with 13,000, 1,000 and 1,000 examples respectively. A single dense layer without any activation is trained and tested on the learned representations for 200 epochs, using Adam optimizer with a learning rate of 0.0025. We use coefficient of determination $R^{2}$ to assess the performance of the single layer network, and obtain $R^2$ = 0.93, 0.64, 0.72, -0.01, 0.85, 0.20 for $\Omega_{\rm m}$, $\sigma_{8}$, $A_{\rm SN1}$, $A_{\rm AGN1}$, $A_{\rm ASN2}$ and $A_{\rm AGN2}$ respectively. 
Results suggest that the latent space representations of the images still carries relevant information for the feed-forward neural network to recover some parameters to a relatively good accuracy ($R^{2} > 0.7$). The recovery of the matter density $\Omega_{\rm m}$ corresponds to the highest accuracy $R^{2} = 0.93$, which is consistent with the most apparent clustering of data points of similar $\Omega_{\rm m}$ values. The challenge related to the predictions of the AGN feedback parameters was also highlighted in \cite{villaescusa2021multifield}. 

From sampling in the latent space, the model is capable of generating new examples which are similar to the training dataset using the decoder. 
Field images of a real instance from the test set are qualitatively compared with those of a new instance generated by the decoder in Figure \ref{fig:fake images}. Visually, the quality of the generated images, exhibiting the topological features like voids, filament, halos and bubbles, is comparable with that of the data. However the fake images tend to be slightly blurry, which points toward the fact that capturing the small scale features is challenging to our model. 
To investigate the statistical properties of the generated 
instances, we compute: the average ($\bar{P}^{\rm XX}(k)$) 
and standard deviation ($\sigma_{{P}^{\rm XX}}(k)$) of auto-power spectrum of each channel of all images in the test set, and both average  ($\bar{P}^{\rm XY}(k)$) and standard deviation ($\sigma_{{P}^{\rm XY}}(k)$) of the cross-power spectrum between channels of all the test instances. We plot $\bar{P}^{\rm XX}(k)$ 
 for each field at the bottom row of Figure \ref{fig:power spectra}. The real and fake power spectra are in good agreement down to $k\sim 4h/{\rm Mpc}$. This comes to corroborate the fact that the generated images are a bit blurred, causing their power spectra to be lower than those of the real images on smaller scales. The  consistency between the average cross-power spectra between channels of the real and generated instances up to $k \sim 5h/{\rm Mpc}$ (Figure \ref{fig:power spectra} top row) indicates that the channels of the generated samples  are consistent with one another, i.e. the three field images describe the same physical region. The clustering properties of the fake maps demonstrate that the generative model is capable of producing maps that mimic the training data.

\section{Conclusion}
We have demonstrated in this work that meaningful latent space representations of CAMELS multifield dataset can be learned using Very Deep Variational AutoEncoder (VDVAE), which is a generative model. The latent codes, which still carry useful information, can be used for some downstream task such as recovering the cosmological and some astrophysical parameters of the maps to a relatively good accuracy. The new maps generated by the model are in good agreement with the data up to a level of high frequency, as evidenced by the consistency between the cross/auto-power spectra of fake and real images for a wavenumber $k \le 5h/{\rm Mpc}$. This limitation of the model can be addressed by further fine-tuning the decay of the KL-divergence part of the loss function, or increasing the model capacity or the combination of the two. For future work, a normalizing flow based model can be trained on the latent codes. This will provide the likelihood of the inputs, which is very important for parameter inference or out-of-distribution task. In order to investigate the robustness of the learned representations, the trained model can be tested on different datasets, such as SIMBA. 

\begin{ack}
SA acknowledges financial support from the South African Radio Astronomy Observatory (SARAO). SH acknowledges support for Program number HST-HF2-51507 provided by NASA through a grant from the Space Telescope Science Institute, which is operated by the Association of Universities for Research in Astronomy, incorporated, under NASA contract NAS5-26555. SH also acknowledges support through NYU Postdoctoral Research and Professional Development Support Grants and Simons Foundation.
\end{ack}

\clearpage
\bibliographystyle{unsrtnat}
\bibliography{references}

\begin{thebibliography}{22}
\providecommand{\natexlab}[1]{#1}
\providecommand{\url}[1]{\texttt{#1}}
\expandafter\ifx\csname urlstyle\endcsname\relax
  \providecommand{\doi}[1]{doi: #1}\else
  \providecommand{\doi}{doi: \begingroup \urlstyle{rm}\Url}\fi

\bibitem[Aghamousa et~al.(2016)Aghamousa, Aguilar, Ahlen, Alam, Allen, Prieto,
  Annis, Bailey, Balland, Ballester, et~al.]{aghamousa2016desi}
Amir Aghamousa, Jessica Aguilar, Steve Ahlen, Shadab Alam, Lori~E Allen,
  Carlos~Allende Prieto, James Annis, Stephen Bailey, Christophe Balland, Otger
  Ballester, et~al.
\newblock The desi experiment part i: science, targeting, and survey design.
\newblock \emph{arXiv:1611.00036}, 2016.

\bibitem[Dewdney et~al.(2009)Dewdney, Hall, Schilizzi, and
  Lazio]{dewdney2009square}
Peter~E Dewdney, Peter~J Hall, Richard~T Schilizzi, and T~Joseph~LW Lazio.
\newblock The square kilometre array.
\newblock \emph{Proceedings of the IEEE}, 97\penalty0 (8):\penalty0 1482--1496,
  2009.

\bibitem[Weltman et~al.(2020)Weltman, Bull, Camera, Kelley, Padmanabhan,
  Pritchard, Raccanelli, Riemer-S{\o}rensen, Shao, Andrianomena,
  et~al.]{weltman2020fundamental}
Amanda Weltman, P~Bull, S~Camera, K~Kelley, H~Padmanabhan, J~Pritchard,
  A~Raccanelli, S~Riemer-S{\o}rensen, L~Shao, S~Andrianomena, et~al.
\newblock Fundamental physics with the square kilometre array.
\newblock \emph{Publications of the Astronomical Society of Australia},
  37:\penalty0 e002, 2020.

\bibitem[Newburgh et~al.(2016)Newburgh, Bandura, Bucher, Chang, Chiang, Cliche,
  Dav{\'e}, Dobbs, Clarkson, Ganga, et~al.]{newburgh2016hirax}
LB~Newburgh, K~Bandura, MA~Bucher, T-C Chang, HC~Chiang, JF~Cliche, R~Dav{\'e},
  M~Dobbs, C~Clarkson, KM~Ganga, et~al.
\newblock Hirax: a probe of dark energy and radio transients.
\newblock In \emph{Ground-based and Airborne Telescopes VI}, volume 9906, pages
  2039--2049. SPIE, 2016.

\bibitem[Gupta et~al.(2018)Gupta, Matilla, Hsu, and Haiman]{gupta2018non}
Arushi Gupta, Jos{\'e} Manuel~Zorrilla Matilla, Daniel Hsu, and Zolt{\'a}n
  Haiman.
\newblock Non-gaussian information from weak lensing data via deep learning.
\newblock \emph{Physical Review D}, 97\penalty0 (10):\penalty0 103515, 2018.

\bibitem[Ribli et~al.(2019)Ribli, Pataki, and Csabai]{ribli2019improved}
Dezs{\H{o}} Ribli, B{\'a}lint~{\'A}rmin Pataki, and Istv{\'a}n Csabai.
\newblock An improved cosmological parameter inference scheme motivated by deep
  learning.
\newblock \emph{Nature Astronomy}, 3\penalty0 (1):\penalty0 93--98, 2019.

\bibitem[Fluri et~al.(2019)Fluri, Kacprzak, Lucchi, Refregier, Amara, Hofmann,
  and Schneider]{fluri2019cosmological}
Janis Fluri, Tomasz Kacprzak, Aurelien Lucchi, Alexandre Refregier, Adam Amara,
  Thomas Hofmann, and Aurel Schneider.
\newblock Cosmological constraints with deep learning from kids-450 weak
  lensing maps.
\newblock \emph{Physical Review D}, 100\penalty0 (6):\penalty0 063514, 2019.

\bibitem[Villaescusa-Navarro et~al.(2021{\natexlab{a}})Villaescusa-Navarro,
  Angl{\'e}s-Alc{\'a}zar, Genel, Spergel, Li, Wandelt, Nicola, Thiele, Hassan,
  Matilla, et~al.]{villaescusa2021multifield}
Francisco Villaescusa-Navarro, Daniel Angl{\'e}s-Alc{\'a}zar, Shy Genel,
  David~N Spergel, Yin Li, Benjamin Wandelt, Andrina Nicola, Leander Thiele,
  Sultan Hassan, Jose Manuel~Zorrilla Matilla, et~al.
\newblock Multifield cosmology with artificial intelligence.
\newblock 2021{\natexlab{a}}.

\bibitem[Andrianomena and Hassan(2023)]{andrianomena2022predictive}
Sambatra Andrianomena and Sultan Hassan.
\newblock Predictive uncertainty on astrophysics recovery from multifield
  cosmology.
\newblock \emph{Journal of Cosmology and Astroparticle Physics}, 2023\penalty0
  (06):\penalty0 051, 2023.

\bibitem[Hassan et~al.(2020)Hassan, Andrianomena, and
  Doughty]{hassan2020constraining}
Sultan Hassan, Sambatra Andrianomena, and Caitlin Doughty.
\newblock Constraining the astrophysics and cosmology from 21 cm tomography
  using deep learning with the ska.
\newblock \emph{Monthly Notices of the Royal Astronomical Society},
  494\penalty0 (4):\penalty0 5761--5774, 2020.

\bibitem[Chen et~al.(2020)Chen, Kornblith, Norouzi, and Hinton]{chen2020simple}
Ting Chen, Simon Kornblith, Mohammad Norouzi, and Geoffrey Hinton.
\newblock A simple framework for contrastive learning of visual
  representations.
\newblock In \emph{International conference on machine learning}, pages
  1597--1607. PMLR, 2020.

\bibitem[Grill et~al.(2020)Grill, Strub, Altch{\'e}, Tallec, Richemond,
  Buchatskaya, Doersch, Avila~Pires, Guo, Gheshlaghi~Azar,
  et~al.]{grill2020bootstrap}
Jean-Bastien Grill, Florian Strub, Florent Altch{\'e}, Corentin Tallec, Pierre
  Richemond, Elena Buchatskaya, Carl Doersch, Bernardo Avila~Pires, Zhaohan
  Guo, Mohammad Gheshlaghi~Azar, et~al.
\newblock Bootstrap your own latent-a new approach to self-supervised learning.
\newblock \emph{Advances in neural information processing systems},
  33:\penalty0 21271--21284, 2020.

\bibitem[He et~al.(2020)He, Fan, Wu, Xie, and Girshick]{he2020momentum}
Kaiming He, Haoqi Fan, Yuxin Wu, Saining Xie, and Ross Girshick.
\newblock Momentum contrast for unsupervised visual representation learning.
\newblock In \emph{Proceedings of the IEEE/CVF conference on computer vision
  and pattern recognition}, pages 9729--9738, 2020.

\bibitem[Chen and He(2021)]{chen2021exploring}
Xinlei Chen and Kaiming He.
\newblock Exploring simple siamese representation learning.
\newblock In \emph{Proceedings of the IEEE/CVF conference on computer vision
  and pattern recognition}, pages 15750--15758, 2021.

\bibitem[Tr{\"o}ster et~al.(2019)Tr{\"o}ster, Ferguson, Harnois-D{\'e}raps, and
  McCarthy]{troster2019painting}
Tilman Tr{\"o}ster, Cameron Ferguson, Joachim Harnois-D{\'e}raps, and Ian~G
  McCarthy.
\newblock Painting with baryons: augmenting n-body simulations with gas using
  deep generative models.
\newblock \emph{Monthly Notices of the Royal Astronomical Society: Letters},
  487\penalty0 (1):\penalty0 L24--L29, 2019.

\bibitem[Yi et~al.(2020)Yi, Guo, Fan, Hamann, and Wang]{yi2020cosmo}
Kai Yi, Yi~Guo, Yanan Fan, Jan Hamann, and Yu~Guang Wang.
\newblock Cosmo vae: Variational autoencoder for cmb image inpainting.
\newblock In \emph{2020 International Joint Conference on Neural Networks
  (IJCNN)}, pages 1--7. IEEE, 2020.

\bibitem[Thorne et~al.(2021)Thorne, Knox, and Prabhu]{thorne2021generative}
Ben Thorne, Lloyd Knox, and Karthik Prabhu.
\newblock A generative model of galactic dust emission using variational
  autoencoders.
\newblock \emph{Monthly Notices of the Royal Astronomical Society},
  504\penalty0 (2):\penalty0 2603--2613, 2021.

\bibitem[Horowitz et~al.(2022)Horowitz, Dornfest, Luki{\'c}, and
  Harrington]{horowitz2022hyphy}
Benjamin Horowitz, Max Dornfest, Zarija Luki{\'c}, and Peter Harrington.
\newblock Hyphy: Deep generative conditional posterior mapping of
  hydrodynamical physics.
\newblock \emph{The Astrophysical Journal}, 941\penalty0 (1):\penalty0 42,
  2022.

\bibitem[Villaescusa-Navarro et~al.(2022)Villaescusa-Navarro, Genel,
  Angles-Alcazar, Thiele, Dave, Narayanan, Nicola, Li, Villanueva-Domingo,
  Wandelt, et~al.]{CMD}
Francisco Villaescusa-Navarro, Shy Genel, Daniel Angles-Alcazar, Leander
  Thiele, Romeel Dave, Desika Narayanan, Andrina Nicola, Yin Li, Pablo
  Villanueva-Domingo, Benjamin Wandelt, et~al.
\newblock The camels multifield data set: Learning the universe’s fundamental
  parameters with artificial intelligence.
\newblock \emph{The Astrophysical Journal Supplement Series}, 259\penalty0
  (2):\penalty0 61, 2022.

\bibitem[Villaescusa-Navarro et~al.(2021{\natexlab{b}})Villaescusa-Navarro,
  Angl{\'e}s-Alc{\'a}zar, Genel, Spergel, Somerville, Dave, Pillepich,
  Hernquist, Nelson, Torrey, et~al.]{villaescusa2021camels}
Francisco Villaescusa-Navarro, Daniel Angl{\'e}s-Alc{\'a}zar, Shy Genel,
  David~N Spergel, Rachel~S Somerville, Romeel Dave, Annalisa Pillepich, Lars
  Hernquist, Dylan Nelson, Paul Torrey, et~al.
\newblock The camels project: Cosmology and astrophysics with machine-learning
  simulations.
\newblock \emph{The Astrophysical Journal}, 915\penalty0 (1):\penalty0 71,
  2021{\natexlab{b}}.

\bibitem[Child(2020)]{child2020very}
Rewon Child.
\newblock Very deep vaes generalize autoregressive models and can outperform
  them on images.
\newblock \emph{arXiv:2011.10650}, 2020.

\bibitem[Kingma and Welling(2013)]{kingma2013auto}
Diederik~P Kingma and Max Welling.
\newblock Auto-encoding variational bayes.
\newblock \emph{arXiv preprint arXiv:1312.6114}, 2013.

\end{thebibliography}


\end{document}